\documentclass[lettersize,journal]{IEEEtran}
\usepackage{amsmath,amsfonts}
\usepackage{algorithmic}
\usepackage{algorithm}
\usepackage{array}
\usepackage[caption=false,font=normalsize,labelfont=sf,textfont=sf]{subfig}
\usepackage{textcomp}
\usepackage{stfloats}
\usepackage{url}
\usepackage{verbatim}
\usepackage{graphicx}
\usepackage{booktabs}
\usepackage{cite}
\usepackage{tabularx}
\usepackage{multirow}
\usepackage{wasysym} 
\usepackage{orcidlink}   
\usepackage{marvosym}    
\usepackage{marvosym}

\begin{document}

\title{Two-Stage Multi-modal Learning for Dynamic Audio-Visual Navigation: From Physical Alignment to Scene Aggregation}

\author{
Jia Li$^{1,2,3}$, and Yinfeng Yu$^{1,2,3}$$^{,\mbox{\Letter}}$$^{,\orcidlink{0000-0003-3089-4140}}$, ~\IEEEmembership{Member,~IEEE} %
\thanks{$^{\mbox{\Letter}}$Yinfeng Yu is the corresponding author(E-mail: yuyinfeng@xju.edu.cn).}%
\\

$^1$School of Computer Science and Technology, Xinjiang University, Urumqi 830017, China\\
$^2$Joint Research Laboratory for Embodied Intelligence, Xinjiang University\\
$^3$Joint International Research Laboratory of Silk Road Multilingual Cognitive Computing, Xinjiang University\\

}



\maketitle

\begin{abstract}
Embodied Audio-Visual Navigation (AVN) requires an agent to locate sound targets by jointly leveraging visual and auditory cues. While prior work has achieved notable success with static sound sources, real-world scenarios often involve moving targets, giving rise to the more challenging problem of \textbf{Dynamic Audio-Visual Navigation (D-AVN)}. Existing approaches largely rely on semantic auditory features, which limits generalization to unheard sound categories and hinders fine-grained spatial reasoning.
To address these challenges, we propose a novel \textbf{Two-Stage Multi-modal Learning} framework for D-AVN. In the Physical Alignment stage, the proposed \textbf{Intensity-Calibrated Geometry Encoder (ICGE)} performs physics-informed alignment of sound intensity with visual geometry to capture category-agnostic spatial correspondence. In the Scene Aggregation stage, a Transformer-based \textbf{Cross-Modal Scene Aggregator (CMSA)} dynamically integrates multi-modal features to construct a robust state representation for policy learning. Extensive experiments on Replica and Matterport3D demonstrate that our method achieves state-of-the-art success rates, while significantly improving generalization and navigation efficiency in complex environments and under unheard sound scenarios.
\end{abstract}

\begin{IEEEkeywords}
Audio-Visual Navigation, Reinforcement Learning, Policy Learning, Cross-Modal Fusion.
\end{IEEEkeywords}

\section{Introduction}
\label{intro}

\begin{figure*}[htbp]
\centering
\includegraphics[scale=0.68]{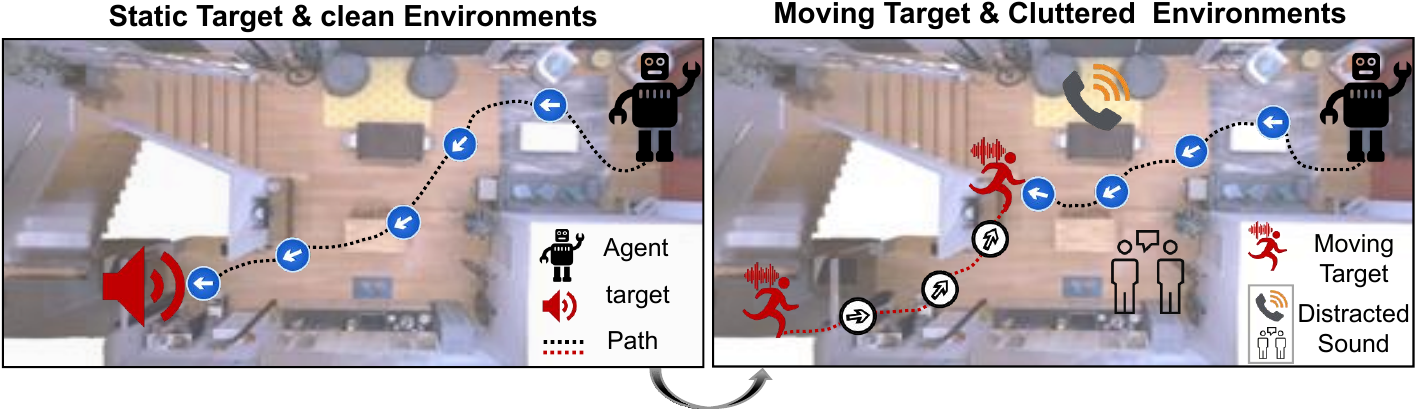}
\caption{Task comparison. Left: static sound source—target is fixed in a clean environment. Right: moving sound source—target moves with interference, creating a dynamic and more challenging task.}
\label{figure1}
\end{figure*}

\IEEEPARstart{E}mbodied AI~\cite{cao2024vnet,zhang2024nonlinear,mattursun2024bss,wang2025modality} research~\cite{liu2025embodied,qin2026audio,yang2026beyond} aims to enable intelligent agents to accomplish complex tasks through continuous interaction with the physical world. Among its core competencies, autonomous navigation has remained a long-standing and fundamental research problem. Conventional navigation systems predominantly rely on geometric or appearance cues extracted from visual observations or LiDAR measurements~\cite{qin2022survey}. While effective in controlled settings, such unimodal perception is inherently fragile in real-world environments characterized by visual occlusion, poor illumination, or limited visibility. In contrast, sound provides a complementary sensing modality that can propagate around obstacles and convey long-range spatial cues, thereby enriching environmental perception beyond the visual field. Motivated by this observation, Audio-Visual Navigation (AVN)~\cite{chen2020soundspaces,chen2021waypoints,chaplot2020learning} has emerged as a promising paradigm, demonstrating strong performance in navigating toward static sound sources by jointly leveraging auditory and visual signals.

However, many real-world navigation scenarios are inherently dynamic, involving moving targets such as walking humans or mobile devices. In such cases, the static-target assumption underlying most AVN methods no longer holds. This gives rise to the task of Dynamic Audio-Visual Navigation (D-AVN)~\cite{younes2023catch,Yu_2022_BMVC}, where the agent must continuously perceive and reason over time-varying audio-visual observations, infer the motion of the sound source, and adapt its navigation strategy accordingly. As illustrated in Figure~\ref{figure1}, D-AVN is substantially more challenging than its static counterpart, requiring the agent to handle transient acoustic signals, spatially complex environments, and persistent interference, all while performing real-time decision making.

Despite recent progress, existing D-AVN approaches suffer from two fundamental limitations. 
First, many methods rely heavily on sound semantic features learned from training data, such as category-specific spectral patterns. 
Although effective for known sound categories, this semantic-centric paradigm generalizes poorly to unheard or out-of-distribution sounds, as it overlooks category-agnostic cues consistent with acoustic propagation principles, particularly the relationship between acoustic intensity and environmental geometry. 
Second, prevailing multi-modal fusion strategies often adopt simplistic late-fusion schemes, such as feature concatenation or coarse attention over global embeddings. 
These static fusion mechanisms fail to preserve fine-grained spatial correspondence between auditory and visual cues, thereby limiting accurate state reasoning and robust planning in dynamic environments.

To address these challenges, we propose a \textbf{Two-Stage Multi-modal Learning} framework for Dynamic Audio-Visual Navigation, which decomposes multi-modal reasoning into physics-informed physical alignment and scene aggregation. 
In the Physical Alignment stage, we introduce the \textbf{Intensity-Calibrated Geometry Encoder (ICGE)}, which performs pixel-level fusion of sound intensity and depth geometry to learn category-agnostic intensity--geometry co-occurrence. 
In the scene aggregation stage, we design the \textbf{Cross-Modal Scene Aggregator (CMSA)} based on a Transformer architecture, enabling dynamic, fine-grained interaction among spatially preserved multi-modal features and overcoming the limitations of static late-fusion schemes.

We evaluate the proposed framework in photorealistic 3D environments, including Replica and Matterport3D~\cite{Matterport3D}. Extensive experimental results demonstrate that our approach learns robust and generalizable navigation policies for the D-AVN task. In particular, our method achieves substantial improvements in success rate when generalizing to unheard sound categories, validating the effectiveness of the proposed two-stage paradigm.

\textbf{Our main contributions are summarized as follows:}
\begin{itemize}
\item We propose a novel Two-Stage Multi-modal Learning framework for D-AVN, including an Intensity-Calibrated Geometry Encoder (ICGE) that performs physics-informed physical alignment via category-agnostic intensity--geometry fusion.
\item We design a Transformer-based Cross-Modal Scene Aggregator (CMSA) to realize dynamic fine-grained cross-modal interaction while maintaining spatial structure.
\item Extensive experiments on Replica and Matterport3D verify that our method achieves new SOTA D-AVN performance and improves generalization to unheard sound categories.
\end{itemize}

\section{Related Work}
\label{relate}

\subsection{Embodied Audio-Visual Navigation}
Embodied navigation~\cite{qin2026audio,yu2025dope} studies how an agent autonomously moves and accomplishes goals through continuous interaction with a 3D environment. Early approaches primarily relied on geometric and visual cues, while subsequent research has explored more semantic-driven navigation tasks, such as object-goal navigation~\cite{putta2024embodiment,chen2023not,wang2025curiosity} and Vision-Language Navigation (VLN)~\cite{wu2024vision,song2025towards}. Although these methods have achieved substantial progress, they predominantly depend on visual or linguistic modalities, which limits robustness in complex and dynamic environments. To overcome this, recent studies have integrated supplementary modalities, notably sound—a highly informative cue owing to its long-range and occlusion-penetrating propagation~\cite{yu2023measuring,wen2025zero}.

The availability of photorealistic simulators with physically accurate acoustic rendering, such as Habitat~\cite{szot2021habitat,savva2019habitat}, has further advanced research on Audio-Visual Navigation (AVN)~\cite{chen2020soundspaces,gan2020look}. Early AVN studies primarily focused on static sound source navigation, where the target remains fixed and agents learn to approach it by fusing binaural audio with visual observations~\cite{wu2024embodied,wang2023learning,zhao2025audio,shi2025towards,chen2023omnidirectional}. These methods demonstrated the effectiveness of auditory cues as a complement to vision, particularly under visually challenging conditions.

More recently, Dynamic Audio-Visual Navigation (D-AVN) has been introduced to handle scenarios with moving sound sources. CMHM~\cite{younes2023catch} formally defined the D-AVN task and established standardized evaluation protocols with dynamic targets and acoustic interference. Building upon this benchmark, FSAAVN~\cite{Yu_2022_BMVC} incorporated self-attention mechanisms to improve the flexibility of audio-visual cue utilization. MCAHAVN~\cite{HE202669} further combines explicit heatmap grounding with cyclic cross-attention fusion for chasing moving sources in complex acoustics. Despite these advances, existing D-AVN methods largely rely on a semantic-centric learning paradigm, which limits their generalization capability when encountering unheard sound categories or complex acoustic dynamics.

\subsection{Multi-modal Feature Fusion}
Multi-modal feature fusion plays a critical role in AVN and other embodied tasks~\cite{han2025multimodal,yu2025dgfnet}. A widely adopted strategy follows an encoding-concatenation paradigm, where modality-specific encoders extract features that are subsequently fused via simple concatenation or weighted aggregation before policy learning~\cite{yusound,chen2021waypoints}. Representative methods such as AV-WaN~\cite{chen2021waypoints} employ late fusion over encoded features, while SAVi~\cite{chen2021semantic} and FSAAVN~\cite{Yu_2022_BMVC} further introduce attention mechanisms on pooled global or temporal embeddings. Although computationally efficient, these strategies discard fine-grained spatial structure and therefore provide limited support for precise audio--geometry correspondence in dynamic environments.

More recent studies have explored Transformer-based fusion for richer cross-modal interaction~\cite{shi2025towards,fu2025fsdenet,yu2025dynamic}. For example, IRCAM-AVN~\cite{zhang2025iterative} jointly performs cross-modal fusion and temporal modeling through iterative cross-attention, while MCAHAVN~\cite{HE202669} adopts cyclic cross-attention over heatmap representations. Despite improved feature interaction, these methods mainly operate on high-level semantic representations and do not explicitly aggregate category-agnostic physical cues that preserve spatial correspondence between acoustic intensity and scene geometry. These observations motivate our hierarchical multi-modal reasoning framework, where ICGE first establishes physics-informed physical alignment and CMSA subsequently performs adaptive scene-level aggregation over spatially preserved multi-modal features for Dynamic Audio-Visual Navigation.

\section{Problem Statement}
\label{problem statement}
We study the task of Dynamic Audio-Visual Navigation (D-AVN), where an embodied agent is required to track and reach a continuously \textbf{moving sound source} in an unknown 3D indoor environment~\cite{younes2023catch}. The agent relies solely on sequential ego-centric observations to infer environmental structure, target dynamics, and navigability, and must continuously update its internal state and navigation policy in response to the time-varying target location. Experiments are conducted in high-fidelity simulated environments reconstructed from real-world scenes, including Replica and Matterport3D~\cite{Matterport3D}. We adopt the Habitat framework~\cite{savva2019habitat} together with the SoundSpaces acoustic simulator~\cite{chen2020soundspaces}, which provides physically accurate binaural audio rendering via precomputed Room Impulse Responses (RIRs). By convolving RIRs with sound signals and transforming the resulting audio using STFT, time-frequency spectrograms are obtained and used as the agent’s auditory input~\cite{chen2021waypoints,younes2023catch}.

Following prior D-AVN settings~\cite{younes2023catch}, the sound source continuously moves toward randomly sampled reachable goal locations and resamples upon arrival. The source travels along the shortest navigable path with a speed slightly lower than that of the agent, ensuring task solvability. Unlike static AVN scenarios, the target’s continuously changing position introduces strong temporal dependency, requiring the navigation policy to exhibit both adaptability and predictive capability. To further evaluate robustness and generalization, a challenging complex acoustic setup is adopted~\cite{younes2023catch}, which includes: (1) \textbf{Interference Sources}, where distractor sounds with randomized positions and audibility are introduced; (2) \textbf{Secondary Target Signals}, where additional sound categories may overlap at the target location; and (3) \textbf{Input Augmentation}, where time or frequency masking is applied to the spectrograms. These perturbations significantly increase task difficulty by corrupting semantic cues and acoustic clarity~\cite{younes2023catch}.

The D-AVN task is further formalized as a sequential decision-making problem under the reinforcement learning framework. At each navigation step $t$, the agent receives a depth visual observation $V_t \in \mathbb{R}^{128 \times 128 \times 1}$ and a binaural audio spectrogram $B_t$ (with dimensions $65 \times 26 \times 2$ for Matterport3D and $65 \times 69 \times 2$ for Replica)~\cite{chen2022soundspaces}. 
Given the current observation $o_t = (V_t, B_t)$ and the previous recurrent state $s_{t-1}$, the agent samples an action according to $a_t \sim \pi(a_t \mid o_t, s_{t-1})$.
The discrete action space consists of four actions: Move Forward, Turn Left, Turn Right, and Stop, where the Stop action succeeds only when the agent is within an acceptable tolerance of the sound source’s navigable node~\cite{younes2023catch}. 
The agent’s objective is to learn a policy that maximizes the expected discounted cumulative reward, formulated as $\max_{\pi} \; \mathbb{E}_{\pi} \left[ \sum_{t=1}^{T} \gamma^{t} \, r(s_{t-1}, a_t) \right]$\label{eq2}, where $\gamma$ denotes the discount factor and $r(s_{t-1}, a_t)$ represents the reward function~\cite{younes2023catch}.
Overall, the problem addressed in this study is to design an autonomous agent that, given only ego-centric visual and binaural auditory observations, can robustly track a continuously moving sound source in an unknown and acoustically complex environment while maintaining high navigation efficiency and success rates despite potential interference.

\begin{figure*}[t]
\centering
\includegraphics[scale=0.59]{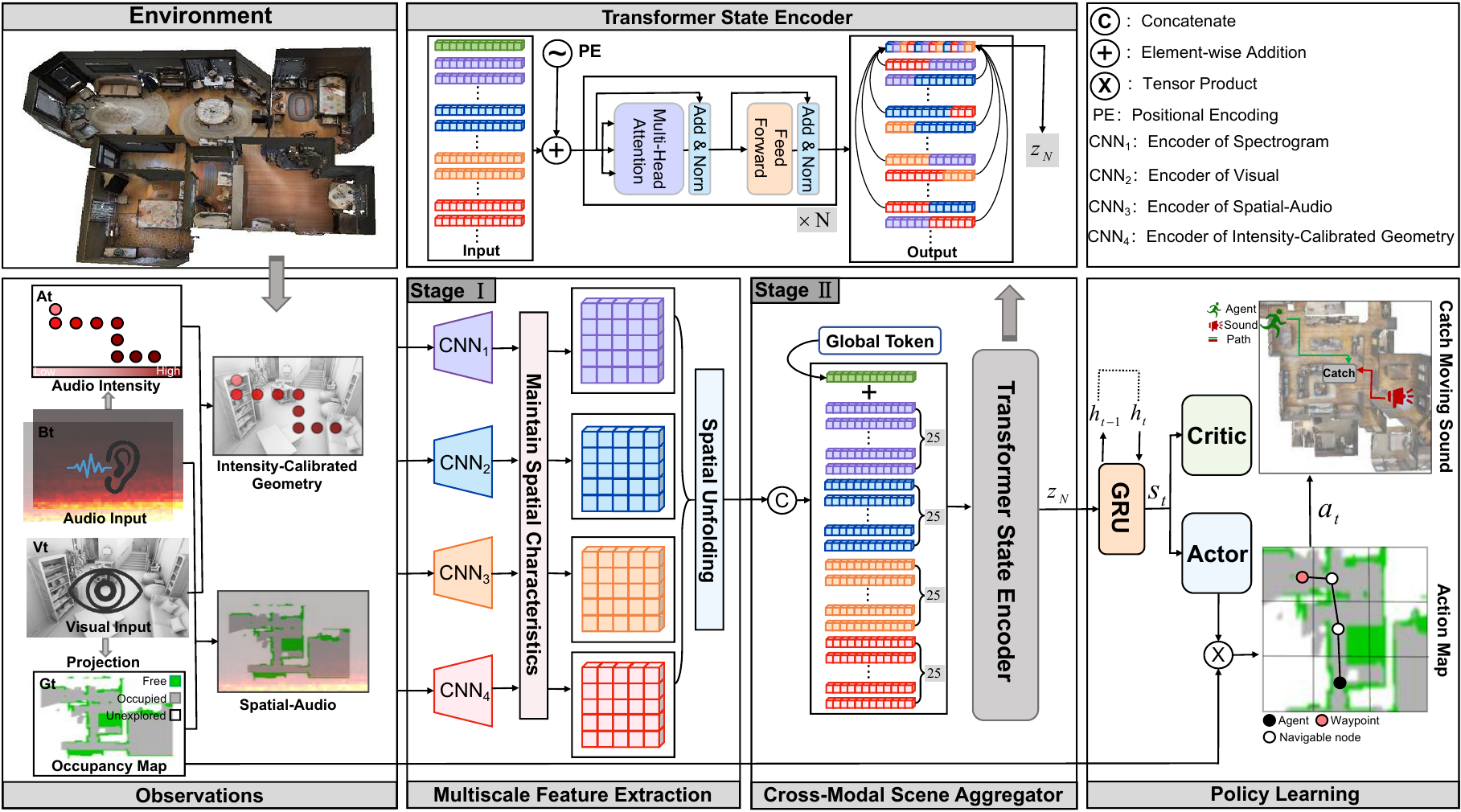}
\caption{Model architecture of the two-stage framework. Observations ($B_t, V_t$) are mapped to $G_t$ and the acoustic memory $A_t$ (this branch in Figure~\ref{figure2}; \emph{not} a global sound field). Stage~I extracts spatial features via four encoders, where ICGE performs physics-informed physical alignment via intensity--geometry fusion. Stage~II aggregates spatial tokens with CMSA to produce $\mathbf{z}_N$. A GRU and PPO-based Actor-Critic complete policy learning.}
\label{figure2}
\end{figure*}

\section{Methods}

To achieve stable tracking and navigation under dynamic sound-source conditions, we propose a Two-Stage Multi-modal Learning framework. The core of our approach is to first establish physics-informed physical alignment between disparate modalities and then perform Scene Aggregation for robust decision-making. As illustrated in Figure~\ref{figure2}, the framework consists of: 
1) Stage I: Physical Alignment, where we extract multi-scale features and perform physics-informed alignment of acoustic intensity with geometry via ICGE; 
2) Stage II: Scene Aggregation, where the CMSA module dynamically fuses multimodal spatial tokens; 
and 3) Policy Learning, which maps the aggregated state to navigation actions. The details of each part are elaborated in the following subsections.

\subsection{Stage I: Physical Alignment and Feature Extraction}
\label{subsec:feature_extraction}

At each time step $t$, the agent receives a multi-modal observation $\mathcal{O}_t = (B_t, V_t)$, where $B_t$ is the binaural spectrogram and $V_t$ is the first-person depth observation~\cite{chen2020soundspaces,chen2021semantic,Chen2024AudioVisual}. To establish a robust spatial correspondence between geometric structure and auditory information, we first process the raw observations into two intermediate map representations, following established methods~\cite{chen2021waypoints}: 1)~Geometric Occupancy Map ($G_t$): An agent-centric occupancy map is continuously constructed and updated from the depth input $V_t$, encoding the scene's geometric structure. 2)~Sound Intensity Map ($A_t$): Sound intensity is computed from the binaural spectrogram $B_t$. The measured intensity is subsequently mapped to the ground-plane locations visited by the agent, yielding a spatial distribution map. This map stores the moving average of the sound intensity values.

Based on $B_t$, $V_t$, $G_t$, and $A_t$, we design four parallel Convolutional Neural Network (CNN) encoders to extract features across different levels and fusion modalities. As shown in Figure~\ref{figure2}, Encoder 1 and 2 extract the basic auditory feature $\mathbf{f}_t^{\text{a}}$ and visual feature $\mathbf{f}_t^{\text{v}}$, respectively. Encoder 3 follows a baseline approach to establish a preliminary association between sound and scene geometry, yielding $\mathbf{f}_t^{\text{sa}}$.

However, a major challenge for the aforementioned encoders is their dependence on sound's semantic spectral features. Because the distribution of these features is strongly correlated with the sound category, the models generalize poorly to unheard sound sources. Inspired by acoustic propagation principles, we leverage the insight that sound intensity provides a category-agnostic cue for locating the source's spatial position~\cite{huang2023discovering}. To learn this cross-category consistency, we introduce Encoder 4: the Intensity-Calibrated Geometry Encoder (ICGE). ICGE learns geometric structures associated with propagation-consistent sound intensity, capturing modality-invariant intensity--geometry co-occurrence.

The ICGE operates by first upsampling the Sound Intensity Map $A_t$ to the same spatial resolution as the depth observation $V_t$ using bilinear interpolation, resulting in $\tilde{A}_t$. This ensures precise spatial alignment between the intensity information and visual pixels. We then concatenate the upsampled intensity map $\tilde{A}_t$ and the depth map $V_t$ along the channel dimension to form the fused input: $\mathbf{I}_t^{\text{fusion}} =[\tilde{A}_t; V_t].$
This fusion input is passed through a dedicated CNN encoder to yield the intensity-calibrated geometric representation $\mathbf{f}_t^{\text{icge}}$. This design implicitly forces the network to learn which geometric structures (e.g., corners, doorways, open spaces) are currently likely to host a high-intensity sound source, transcending mere object category recognition. For instance, a moving source, whether emitting speech, music, or noise, will generate similar "intensity-geometry" co-occurrence patterns as it passes through a doorway. ICGE captures this cross-source, category-invariant intensity--geometry association through end-to-end training, providing a powerful generalization prior. Through this design, the ICGE performs physics-informed representation learning between sound and geometry, providing a category-agnostic spatial prior that is essential for generalizing to unheard sounds. 

Crucially, unlike baseline methods that compress the feature maps into a single vector (via Global Average Pooling or similar techniques)~\cite{younes2023catch, Yu_2022_BMVC}, we explicitly retain the spatial structure of the feature maps from all encoders. Consequently, at time step $t$, we obtain four sets of spatially-rich multi-modal feature maps: $\{\mathbf{f}_t^{\text{a}}, \mathbf{f}_t^{\text{v}}, \mathbf{f}_t^{\text{sa}}, \mathbf{f}_t^{\text{icge}}\}$. Each feature map has spatial dimensions $5 \times 5$ and $C=256$ channels. The specific configuration of the CNN encoders is detailed in Table~\ref{tab:encoder_config}.

\begin{table}[t]
\centering
\caption{CNN encoder configurations.}
\label{tab:encoder_config}
\small
\setlength{\tabcolsep}{8pt}
\begin{tabular}{ll}
\toprule
\textbf{Encoder} & \textbf{Kernel, Stride} \\
\midrule
CNN$_{1}$ & (8, 4), (4, 2), (3, 1) \\
CNN$_{2}$ & (8, 4), (4, 2), (3, 2), (2, 1) \\
CNN$_{3}$ & (8, 4), (4, 2), (3, 2), (3, 2) \\
CNN$_{4}$ & (8, 4), (4, 2), (3, 2), (2, 1) \\
\bottomrule
\end{tabular}
\end{table}

\subsection{Stage II: Cross-Modal Scene Aggregation}

Building upon the multimodal features extracted in Stage I, the Cross-Modal Scene Aggregator (CMSA) acts as the second stage of our learning framework to aggregate these spatially-aligned multi-modal tokens into a unified state representation. The preceding Stage I provides four sets of local feature maps, $\{\mathbf{f}_t^{\text{a}}, \mathbf{f}_t^{\text{v}}, \mathbf{f}_t^{\text{sa}}, \mathbf{f}_t^{\text{icge}}\}$, each with an explicit spatial correspondence. The central challenge now is: how to effectively fuse this heterogeneous multi-modal information into a unified, rich global state representation for robust decision-making~\cite{roy2023multimodal,liu2025embodied}.

Existing works often adopt a Late Concatenation strategy, in which extracted feature maps are globally pooled and compressed into vectors (e.g., $ 512$-dimensional), which are then concatenated and fed into a Recurrent Neural Network (GRU) for temporal decision-making. This strategy suffers from two primary drawbacks: 1)~\textbf{Spatial Information Loss:} Global pooling destroys the fine-grained spatial structure within the feature maps, which is vital for understanding the detailed correspondence between the sound source and the scene geometry. 2)~\textbf{Inflexible Fusion Mode:} Vector concatenation constitutes a static, non-interactive fusion method. It cannot dynamically adjust the importance of different modalities or spatial locations in response to the current context, thereby limiting the model's capacity for complex cross-modal inference.

To overcome these limitations, we propose a deep fusion module based on the Transformer architecture~\cite{SHI2023200}: Cross-Modal Scene Aggregator (CMSA). Its core idea is to treat the multi-modal features as an interactable token sequence, using the self-attention mechanism to achieve fine-grained, content-aware, and dynamic fusion. The procedure is as follows:

First, each $5 \times 5 \times 256$ feature map is flattened to yield a sequence of 25 tokens, each 256-dimensional:
\begin{equation}
\mathbf{X}_t^{(m)} = \text{Flatten}(\mathbf{f}_t^{(m)}), \quad m \in \{\text{a}, \text{v}, \text{sa}, 
\text{icge}\}.
\label{eq4}
\end{equation}
This results in four modal token sequences, totaling $L=100$ spatial tokens. To aggregate the global scene information, we introduce a learnable \textbf{[FUSE]} token, initialized as $\mathbf{z}_0 \in \mathbb{R}^{256}$. This token is prepended to the sequence, serving as a dynamic medium for interaction with all local modal tokens. We then concatenate all tokens and add learnable positional encodings $\mathbf{P} \in \mathbb{R}^{(L+1) \times 256}$ to construct the complete input sequence:
\begin{equation}
\mathbf{H}_0 = [\mathbf{z}_0; \mathbf{X}_t^{(\text{a})}; \mathbf{X}_t^{(\text{v})}; \mathbf{X}_t^{(\text{sa})}; \mathbf{X}_t^{(\text{icge})}] + \mathbf{P}.   
\label{eq5}
\end{equation}
The sequence is then fed into an $N$-layer Transformer encoder. In our experiment, we adopt $N=4$ layers and use $8$ attention heads. Within each layer $l$, the multi-head self-attention mechanism allows each token to adaptively attend to all other tokens in the sequence, thereby enabling deep cross-modal information interaction. This dynamic attention, spanning modalities and space, allows the model to reason about complex joint states such as "the current sound source is likely positioned behind that specific corner."

After $N$ layers, we extract the final \textbf{[FUSE]} token embedding $\mathbf{z}_N$ as the global fused state of the current timestep. Unlike simple concatenation, $\mathbf{z}_N$ is a compact representation obtained by multi-layer fine-grained cross-modal reasoning.

\begin{table}[b]
\centering
\caption{RL training hyperparameters.}
\label{tab:training_config}
\small
\setlength{\tabcolsep}{8pt}
\begin{tabular}{ll}
\toprule
\textbf{Hyperparameter} & \textbf{Value} \\
\midrule
Optimizer & Adam \\
PPO clip & 0.1 \\
PPO epochs & 4 \\
Mini-batches & 1 \\
Value loss coefficient & 0.5 \\
Entropy coefficient & 0.02 \\
Learning rate & $1\times10^{-4}$ \\
GAE $\tau$ / $\gamma$ & 0.95 / 0.99 \\
Hidden size & 256 \\
Steps per rollout & 150 \\
\bottomrule
\end{tabular}
\end{table}

\subsection{Policy Learning}
To enable the agent to capture the temporal evolution of the moving sound source's state in the dynamic environment, the current fused state representation $\mathbf{z}_N$ is first input into a Gated Recurrent Unit (GRU):
\begin{equation}
\mathbf{s}_t, \mathbf{h}_t = \text{GRU}(\mathbf{z}_N, \mathbf{h}_{t-1}).   
\label{eq6}
\end{equation}
Here, $\mathbf{h}_t \in \mathbb{R}^{256}$ is the GRU hidden state, and $\mathbf{s}_t$ is the final state representation, incorporating historical information, used for the current decision-making~\cite{chen2020soundspaces}.

The state vector $\mathbf{s}_t$ is subsequently fed into an Actor-Critic network architecture: the Actor network predicts the probability distribution over all possible actions in the current state, and the Critic network evaluates the value of the current state. The Actor and Critic components are both constructed using fully connected (linear) layers~\cite{zhang2025advancing,Chen2024AudioVisual}. This unified architecture is subsequently optimized end-to-end under the PPO (Proximal Policy Optimization) algorithm ~\cite{schulman2017proximal}. The PPO framework, by constraining the magnitude of policy updates, ensures training stability, which is particularly vital in dynamic, non-stationary moving-sound source environments. Hyperparameters are summarized in Table~\ref{tab:training_config}.

\begin{table*}[t]
    \centering
    \caption{Performance in Clean Environments on Replica and Matterport3D Datasets.}
    \label{table:combined_clean}
    \footnotesize
    \setlength{\tabcolsep}{2.5pt}
    \begin{tabular}{l ccc ccc ccc ccc}
    \toprule
    {\textbf{Model}} & \multicolumn{6}{c}{\textbf{Replica}} & \multicolumn{6}{c}{\textbf{Matterport3D}} \\
    \cmidrule(lr){2-7} \cmidrule(lr){8-13}
    & \multicolumn{3}{c}{Heard} & \multicolumn{3}{c}{Unheard} & \multicolumn{3}{c}{Heard} & \multicolumn{3}{c}{Unheard} \\
    \cmidrule(lr){2-4} \cmidrule(lr){5-7} \cmidrule(lr){8-10} \cmidrule(lr){11-13}
    & DSR$\uparrow$ & DSPL$\uparrow$ & DSNA$\uparrow$ & DSR$\uparrow$ & DSPL$\uparrow$ & DSNA$\uparrow$ & DSR$\uparrow$ & DSPL$\uparrow$ & DSNA$\uparrow$ & DSR$\uparrow$ & DSPL$\uparrow$ & DSNA$\uparrow$ \\
    \midrule
    AV-Nav & 71.0 & 47.2 & 16.3 & 37.4 & 23.4 & 7.8  & 73.0 & 43.4 & 18.4 & 35.5 & 20.9 & 8.4 \\
    AV-WaN & 86.8 & 55.0 & 32.6 & 38.2 & 24.1 & 14.2 & 86.7 & 58.3 & 33.6 & 47.2 & 30.0 & 17.4 \\
    SAVi   & 67.6 & 54.0 & 25.7 & 40.8 & 28.9 & 13.9 & 49.0 & 29.9 & 11.3 & 41.6 & 26.3 & 10.0 \\
    FSAAVN & 80.1 & 52.5 & 28.8 & 39.2 & 28.9 & 15.8 & 83.9 & 47.8 & 24.4 & 43.3 & 27.6 & 18.7 \\
    CMHM   & 90.0 & 62.6 & 34.5 & 49.1 & 29.4 & 16.1 & 94.0 & 63.2 & 35.3 & 57.7 & 37.0 & 21.0 \\
    MCAHAVN & 89.7 & \textbf{67.8} & \textbf{40.2} & 52.6 & \textbf{36.2} & \textbf{24.3} & 92.0 & \textbf{73.7} & \textbf{44.5} & 60.4 & \textbf{49.1} & \textbf{30.8} \\
    \midrule
    \textbf{Ours} & \textbf{93.8} & 67.5 & 37.5 & \textbf{63.5} & 32.1 & 17.8 & \textbf{96.3} & 65.6 & 38.0 & \textbf{69.4} & 39.2 & 22.6 \\
    \bottomrule
    \end{tabular}
\end{table*}

To guide the agent toward efficient and accurate navigation, we employ a reward function consistent with prior works, comprising sparse success rewards and dense guiding signals~\cite{chen2020soundspaces}:
\begin{itemize}
    \item \textbf{Success Reward:} A reward of $+10$ is assigned to the agent upon successfully reaching the target sound source and executing the Stop action.
    \item \textbf{Distance Reward:} For progressive guidance, a reward of $+0.25$ is administered when the agent's Geodesic distance to the target is reduced compared to the preceding step~\cite{zhang2025advancing}.
    \item \textbf{Time Penalty:} A small penalty of $-0.01$ is imposed for every action executed, encouraging navigation efficiency~\cite{chen2021waypoints}.
\end{itemize}

Our structured design yields generalizable navigation policies via multi-level fine-grained audio-visual fusion for dynamic acoustic environments.

\section{Experiments}
We evaluate our framework on dynamic sound environments via experiments, result analysis and ablation studies.

\subsection{Experimental Setup}
We evaluate navigation of moving sound targets under clean and complex acoustic settings.

To evaluate zero-shot generalization, we follow the protocol in~\cite{chen2020soundspaces}, training exclusively on specific sounds and scene instances while testing on unheard environments. We categorize test episodes into \textbf{Heard Sounds} (target sounds belong to categories heard during training) and \textbf{Unheard Sounds} (target sounds belong to categories not present in the training set), using a $73/11/18$ split for the $102$ sound categories (and interference sounds) across Training/Validation/Testing sets. An episode is successful if the agent reaches the moving target and executes a Stop action within $500$ steps~\cite{Yu_2022_BMVC,zhang2025advancing}.

The experimental performance is quantified using a set of established metrics tailored for D-AVN~\cite{younes2023catch,anderson2018evaluation}:

\begin{itemize}
    \item \textbf{DSR (Dynamic Success Rate):} The metric measures the percentage of instances where the agent effectively tracks and reaches the moving sound source objective~\cite{younes2023catch}, as defined in Equation~(\ref{eq7}).
    \begin{equation}
        \text{DSR} = \frac{1}{N} \sum_{i=1}^{N} S_i.
        \label{eq7}
    \end{equation}
    \item \textbf{DSPL (Dynamic Success weighted by Path Length):} Measures the path efficiency upon successful completion~\cite{yang2025knowledge}, as defined in Equation~(\ref{eq8}). For dynamic targets, the optimal path length ($g_i$) is defined as the shortest geodesic distance from the agent's start location to the nearest point on the target's trajectory where capture was possible~\cite{younes2023catch}.
    \begin{equation}
        \text{DSPL} = \frac{1}{N} \sum_{i=1}^{N} S_i \cdot \frac{g_i}{\max(p_i, g_i)}.
        \label{eq8}
    \end{equation}
    \item \textbf{DSNA (Dynamic Success weighted by Number of Actions):} Provides a finer view of policy efficiency by penalizing unnecessary actions, particularly rotation maneuvers, as defined in Equation~(\ref{eq9}). This metric is weighted by the ratio of the minimum required actions to the total actions taken by the agent~\cite{younes2023catch}.
    \begin{equation}
        \text{DSNA} = \frac{1}{N} \sum_{i=1}^{N} S_i \cdot \frac{L_i^{\text{o}}}{\max(L_i^{\text{a}}, L_i^{\text{o}})}.
        \label{eq9}
    \end{equation}
\end{itemize}
In these formulas, $N$ denotes the total number of episodes, and $S_i$ signifies the success indicator. $p_i$ represents the actual path length taken, and $g_i$ refers to the Oracle Optimal Path Length. $L_i^{\text{a}}$ represents the total number of actions executed by the agent, and $L_i^{\text{o}}$ is the minimum required action count from the start to the optimal capture point~\cite{younes2023catch}. While DSPL, DSR, and DSNA evaluate dynamic tracking, the standard metrics SPL, SR, and SNA refer exclusively to static-target scenarios.

\begin{table*}[t]
    \centering
    \caption{Performance in Complex Environments on Replica and Matterport3D Datasets.}
    \label{table:combined_complex}
    \footnotesize
    \setlength{\tabcolsep}{2.5pt}
    \begin{tabular}{l ccc ccc ccc ccc}
    \toprule
    {\textbf{Model}} & \multicolumn{6}{c}{\textbf{Replica}} & \multicolumn{6}{c}{\textbf{Matterport3D}} \\
    \cmidrule(lr){2-7} \cmidrule(lr){8-13}
    & \multicolumn{3}{c}{Heard} & \multicolumn{3}{c}{Unheard} & \multicolumn{3}{c}{Heard} & \multicolumn{3}{c}{Unheard} \\
    \cmidrule(lr){2-4} \cmidrule(lr){5-7} \cmidrule(lr){8-10} \cmidrule(lr){11-13}
    & DSR$\uparrow$ & DSPL$\uparrow$ & DSNA$\uparrow$ & DSR$\uparrow$ & DSPL$\uparrow$ & DSNA$\uparrow$ & DSR$\uparrow$ & DSPL$\uparrow$ & DSNA$\uparrow$ & DSR$\uparrow$ & DSPL$\uparrow$ & DSNA$\uparrow$ \\
    \midrule
    AV-Nav & 60.4 & 34.0 & 11.5 & 45.4 & 26.0 & 8.7  & 63.2 & 34.4 & 13.4 & 46.2 & 24.6 & 9.3  \\
    AV-WaN & 66.3 & 43.1 & 25.1 & 51.5 & 32.8 & 19.0 & 84.0 & 52.3 & 31.5 & 70.3 & 43.0 & 25.6 \\
    SAVi   & 45.5 & 34.2 & 16.1 & 38.3 & 27.7 & 12.7 & 65.6 & 44.0 & 22.0 & 55.5 & 35.1 & 16.0 \\
    FSAAVN & 63.1 & 37.5 & 18.8 & 49.9 & 29.9 & 17.6 & 70.9 & 39.5 & 19.2 & 62.8 & 30.7 & 20.4 \\
    CMHM   & 63.8 & 38.6 & 21.8 & 58.9 & 32.9 & 18.6 & 92.8 & 59.9 & 33.3 & 84.1 & 52.1 & 30.4 \\
    MCAHAVN & 66.7 & 46.8 & \textbf{30.2} & 58.6 & 36.2 & 24.3 & 91.2 & \textbf{72.7} & \textbf{44.3} & 83.4 & \textbf{65.4} & \textbf{41.6} \\
    \midrule
    \textbf{Ours} & \textbf{87.0} & \textbf{53.7} & 30.1 & \textbf{72.9} & \textbf{43.6} & \textbf{24.4} & \textbf{95.9} & 61.3 & 35.7 & \textbf{84.7} & 53.8 & 30.8 \\
    \bottomrule
    \end{tabular}
\end{table*}

\begin{table}[t]
\centering
\caption{Performance of Training in Complex Environments and Testing in Clean Environments.}
\label{table:4}
\setlength{\tabcolsep}{4pt}
\begin{tabular}{l ccc ccc}
\toprule
\textbf{Model} & \multicolumn{3}{c}{\textbf{Heard}} & \multicolumn{3}{c}{\textbf{Unheard}} \\
\cmidrule(lr){2-4} \cmidrule(lr){5-7}
& DSR$\uparrow$ & DSPL$\uparrow$ & DSNA$\uparrow$ 
& DSR$\uparrow$ & DSPL$\uparrow$ & DSNA$\uparrow$ \\
\midrule
\multicolumn{7}{c}{\textbf{Replica}} \\
\midrule
CMHM & 67.6 & 43.9 & 24.5 & \textbf{67.3} & 33.9 & 19.1 \\
\textbf{Ours} & \textbf{79.2} & \textbf{55.6} & \textbf{32.6} & 64.0 & \textbf{37.6} & \textbf{22.3} \\
\midrule
\multicolumn{7}{c}{\textbf{Matterport3D}} \\
\midrule
CMHM & 91.9 & \textbf{61.6} & \textbf{36.2} & 74.0 & \textbf{45.8} & \textbf{26.0} \\
\textbf{Ours} & \textbf{92.2} & 57.4 & 32.4 & \textbf{80.5} & 44.5 & 25.4 \\
\bottomrule
\end{tabular}
\end{table}

\begin{table}[t]
\centering
\caption{Performance with stationary sound sources in clean environments on Matterport3D.}
\label{table:5}
\footnotesize
\setlength{\tabcolsep}{2.5pt}
\begin{tabular}{l ccc ccc}
\toprule
\textbf{Model} & \multicolumn{3}{c}{\textbf{Heard}} & \multicolumn{3}{c}{\textbf{Unheard}} \\
\cmidrule(lr){2-4} \cmidrule(lr){5-7}
& SR$\uparrow$ & SPL$\uparrow$ & SNA$\uparrow$ 
& SR$\uparrow$ & SPL$\uparrow$ & SNA$\uparrow$ \\
\midrule
CMHM & 82.9 & 61.9 & 46.8 & 55.3 & 42.4 & 31.6 \\
AGSA-AVN & 70.0 & 54.1 & 30.0 & 36.5 & 26.2 & 13.1 \\
IRCAM-AVN & -- & -- & -- & 61.5 & 37.8 & 52.4 \\
DMTF-AVN & -- & -- & -- & \textbf{63.1} & 38.4 & \textbf{54.2} \\
\textbf{Ours} & \textbf{84.3} & \textbf{64.9} & \textbf{48.5} & 62.6 & \textbf{48.4} & 35.0 \\
\bottomrule
\end{tabular}
\end{table}

We benchmark our proposed framework against several existing methods as follows:
\begin{itemize}
    \item \textbf{AV-Nav}~\cite{chen2020soundspaces}: An early perception-driven end-to-end reinforcement learning model that relies on raw audio and visual inputs and does not involve explicit map building.
    \item \textbf{AV-WaN}~\cite{chen2021waypoints}: An AVN method that improves efficiency by predicting intermediate waypoints and incorporating explicit geometric and acoustic maps.
    \item \textbf{SAVi}~\cite{chen2021semantic}: Utilizes a Transformer-based architecture to model the temporal evolution of audio-visual features for robust navigation.
    \item \textbf{FSAAVN}~\cite{Yu_2022_BMVC}: A self-attention framework for chasing a moving audio target.
    \item \textbf{CMHM}~\cite{younes2023catch}: Defines the D-AVN benchmark and fuses audio-visual information in the spatial feature space, serving as a strong comparison baseline.
    \item \textbf{MCAHAVN}~\cite{HE202669}: The most recent dynamic D-AVN baseline; it combines audio-visual heatmap grounding with cyclic cross-attention fusion for moving sources.
\end{itemize}

For static AVN, we additionally compare against the most recent methods---IRCAM-AVN~\cite{zhang2025iterative}, DMTF-AVN~\cite{yu2025dynamic}, and AGSA-AVN~\cite{li2025audio}---under the standard AudioGoal protocol (Table~\ref{table:5}).

\subsection{Results and Analysis}

\subsubsection{Quantitative Evaluation}
We first assess the performance of our proposed framework under ideal acoustic conditions (Clean Environments). Table~\ref{table:combined_clean} summarizes the navigation performance on the Replica and Matterport3D (MP3D) datasets.

\begin{figure}[tb]
\centering
\includegraphics[scale=0.60]{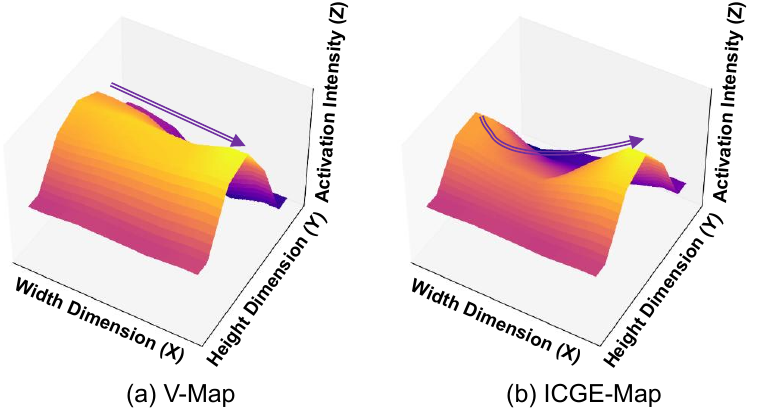}
\caption{Comparison between the pure visual feature map (a) and the feature map of the Intensity-Calibrated Geometry Encoder (b).}
\label{figure3}
\end{figure}

\begin{figure}[b]
\centering
\includegraphics[scale=0.35]{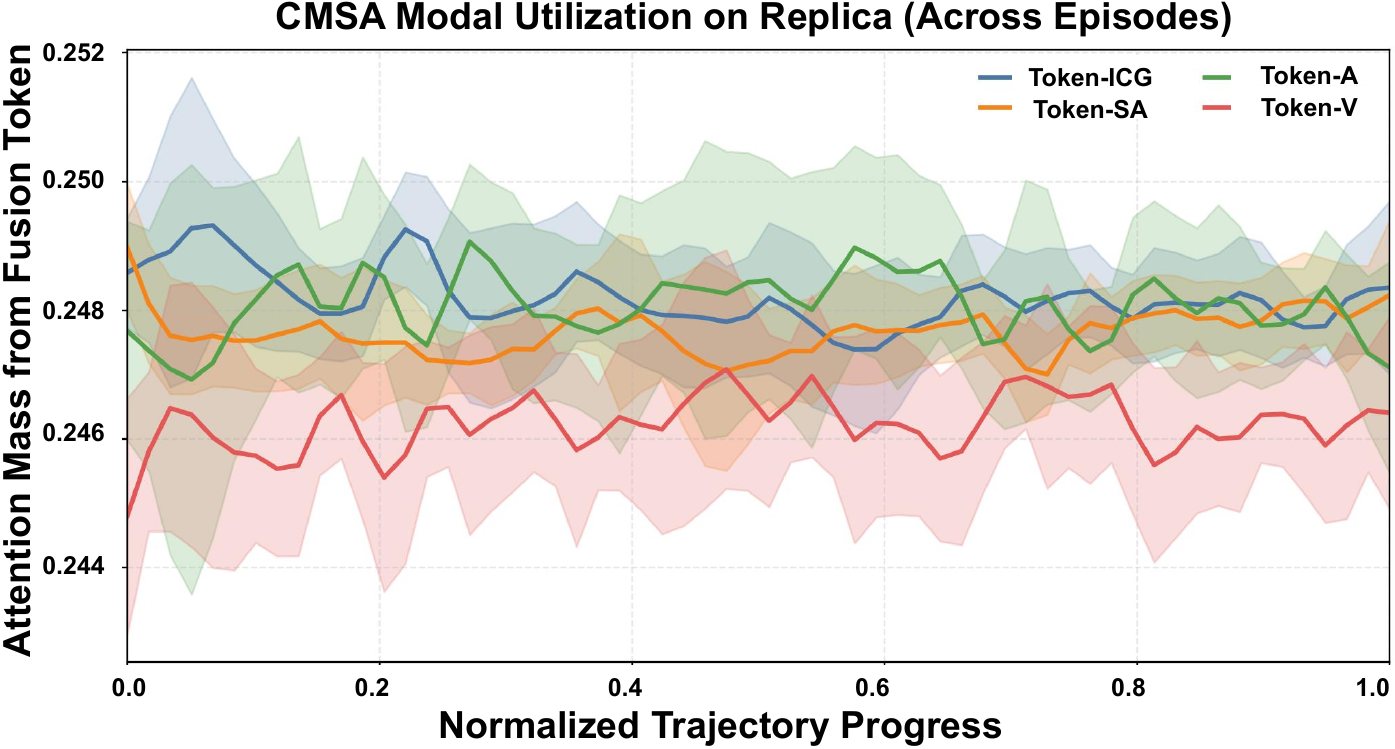}
\caption{Attention distribution across different modalities in the Cross-Modal Scene Aggregator (CMSA) on Replica.}
\label{figure4}
\end{figure}

\begin{figure}[b]
\centering
\includegraphics[scale=0.35]{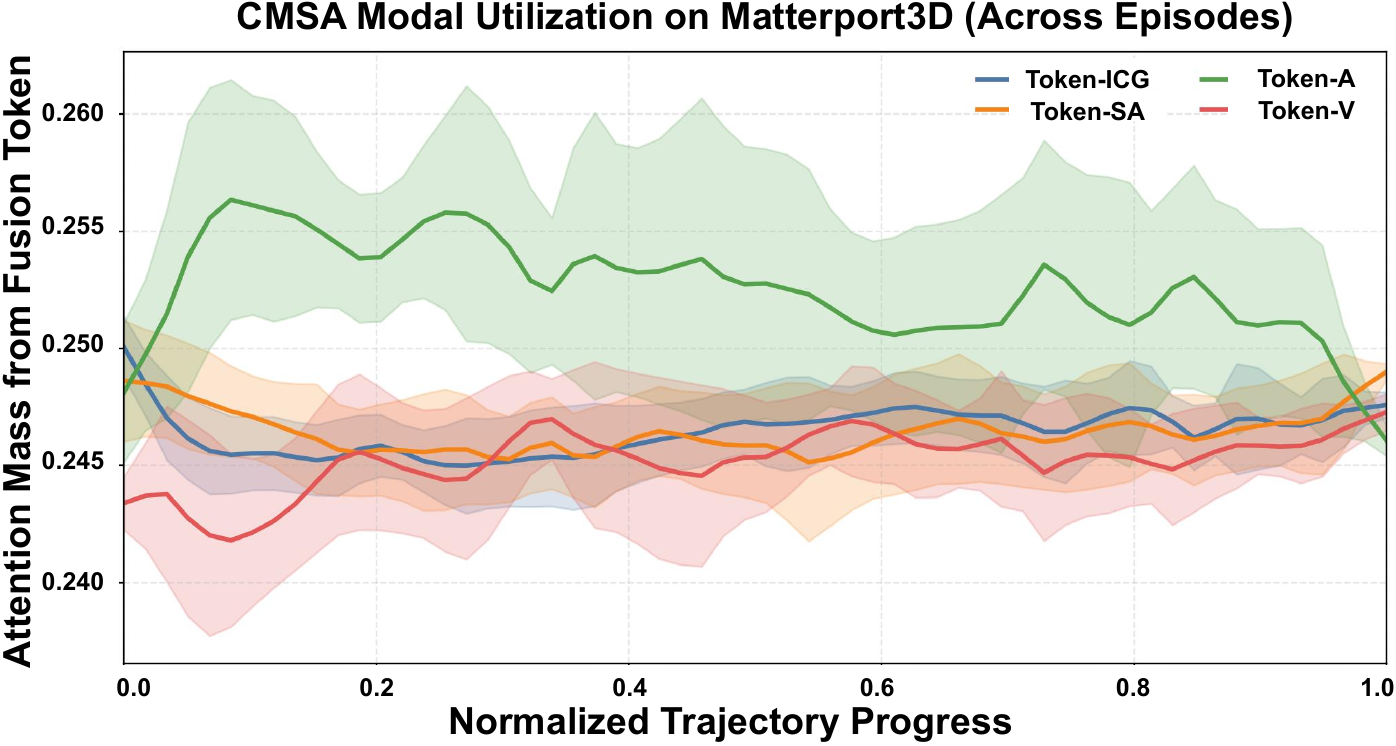}
\caption{Attention distribution across different modalities in the Cross-Modal Scene Aggregator (CMSA) on Matterport3D.}
\label{figure5}
\end{figure}

\begin{figure}[b]
\centering
\includegraphics[scale=0.35]{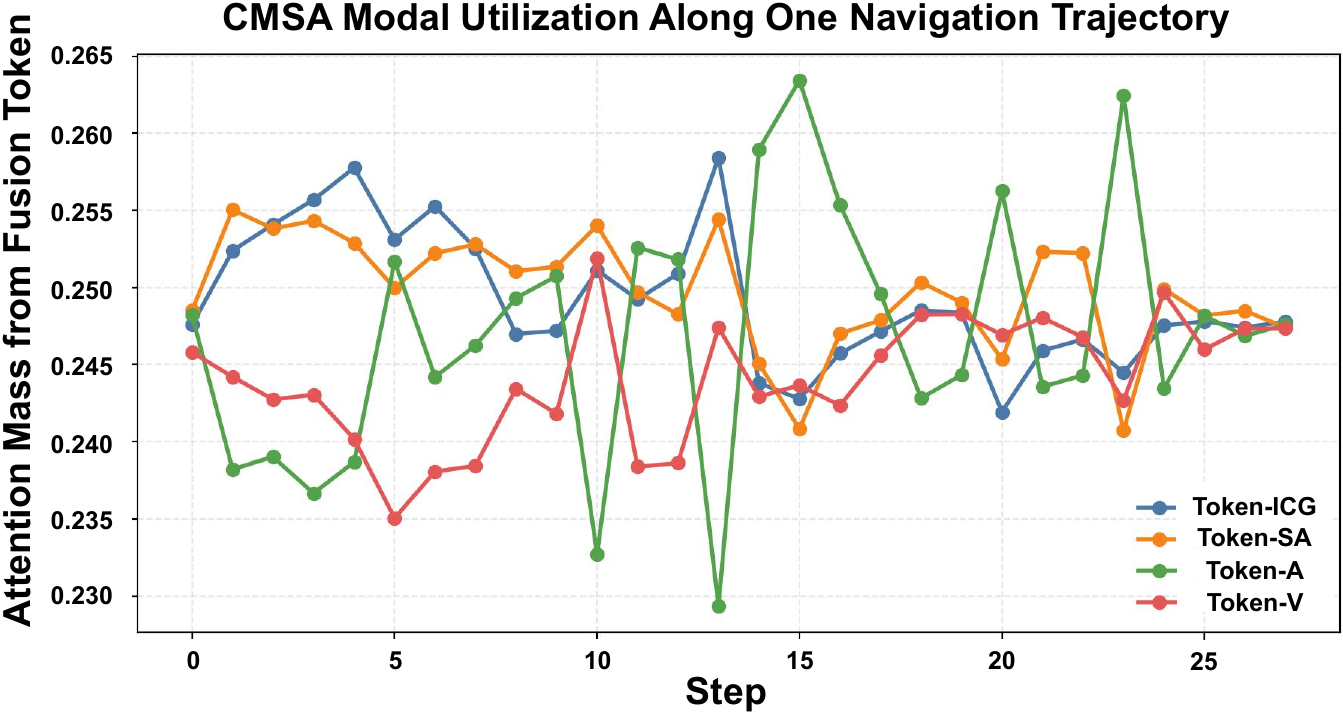}
\caption{Cross-Modal Scene Aggregator modal utilization on Matterport3D through one episode.}
\label{figure6}
\end{figure}

The experimental results demonstrate that our method achieves the strongest \textbf{Unheard} DSR on both datasets (Table~\ref{table:combined_clean}), validating its cross-category generalization.
On Replica, our model reaches $\mathbf{63.5\%}$ Unheard DSR, outperforming the baseline CMHM ($49.1\%$) by $\mathbf{14.4}$ percentage points ($\mathbf{29.3\%}$ relative gain).
On MP3D, Unheard DSR improves from $57.7\%$ (CMHM) to $\mathbf{69.4\%}$, an absolute gain of $\mathbf{11.7}$ percentage points.

These improvements stem from our two-stage paradigm: ICGE establishes physics-informed physical alignment between sound and geometry to ensure robust inference for unheard sounds, while the CMSA employs Transformer-based attention for fine-grained cross-modal reasoning. This combination effectively maintains precise target tracking in dynamic, unfamiliar environments.

Importantly, the framework also attains the highest \textbf{Heard} DSR on both datasets ($\mathbf{93.8\%}$ on Replica and $\mathbf{96.3\%}$ on MP3D). Although DSPL and DSNA are slightly lower than those of MCAHAVN, the proposed method consistently achieves substantially higher success rates, demonstrating a favorable trade-off between navigation efficiency and robustness.

To further assess whether ICGE captures robust physics-informed cues rather than fragile semantic features, we evaluate our framework in Complex Environments (Table~\ref{table:combined_complex}) that introduce reverberation, noise, and interference.
Despite these challenging conditions, our approach consistently achieves the highest DSR while remaining competitive on the remaining metrics. On Replica, the DSR improves from $63.8\%$ to $87.0\%$ and DSPL from $38.6$ to $53.7$ in the Heard setting, while achieving $72.9\%$ DSR and $43.6$ DSPL in the Unheard setting (compared to CMHM's $58.9\%$ and $32.9\%$). On MP3D, although the complex acoustic conditions reduce the overall navigation efficiency (DSPL/DSNA), our approach consistently maintains the highest DSR and remains competitive with the SOTA.

Furthermore, as shown in Table~\ref{table:4}, when trained in complex environments and tested in clean ones, our method maintains its leading performance. This indicates that our framework effectively extracts intrinsic geometric and spatial features from noisy data rather than merely overfitting to specific acoustic patterns.
To further demonstrate the versatility of our framework beyond dynamic scenarios, we evaluate it on the conventional static audio-visual navigation task with stationary sound sources (Table~\ref{table:5}). Despite being designed for D-AVN, our method outperforms a wide range of recent AVN approaches, including AGSA-AVN~\cite{li2025audio}, IRCAM-AVN~\cite{zhang2025iterative}, DMTF-AVN~\cite{yu2025dynamic}, and CMHM. In particular, it achieves the highest SR in both Heard (84.3\%) and Unheard (62.6\%) settings. These results demonstrate that the proposed framework generalizes effectively to static audio-visual navigation and possesses broad applicability.

Overall, these results demonstrate that the proposed framework effectively leverages physics-informed physical alignment and dynamic scene-level aggregation, establishing a new state of the art in terms of navigation success rate (DSR) while substantially improving cross-category generalization and robustness under complex acoustic conditions.

\begin{figure*}[t]
\centering
\includegraphics[scale=0.61]{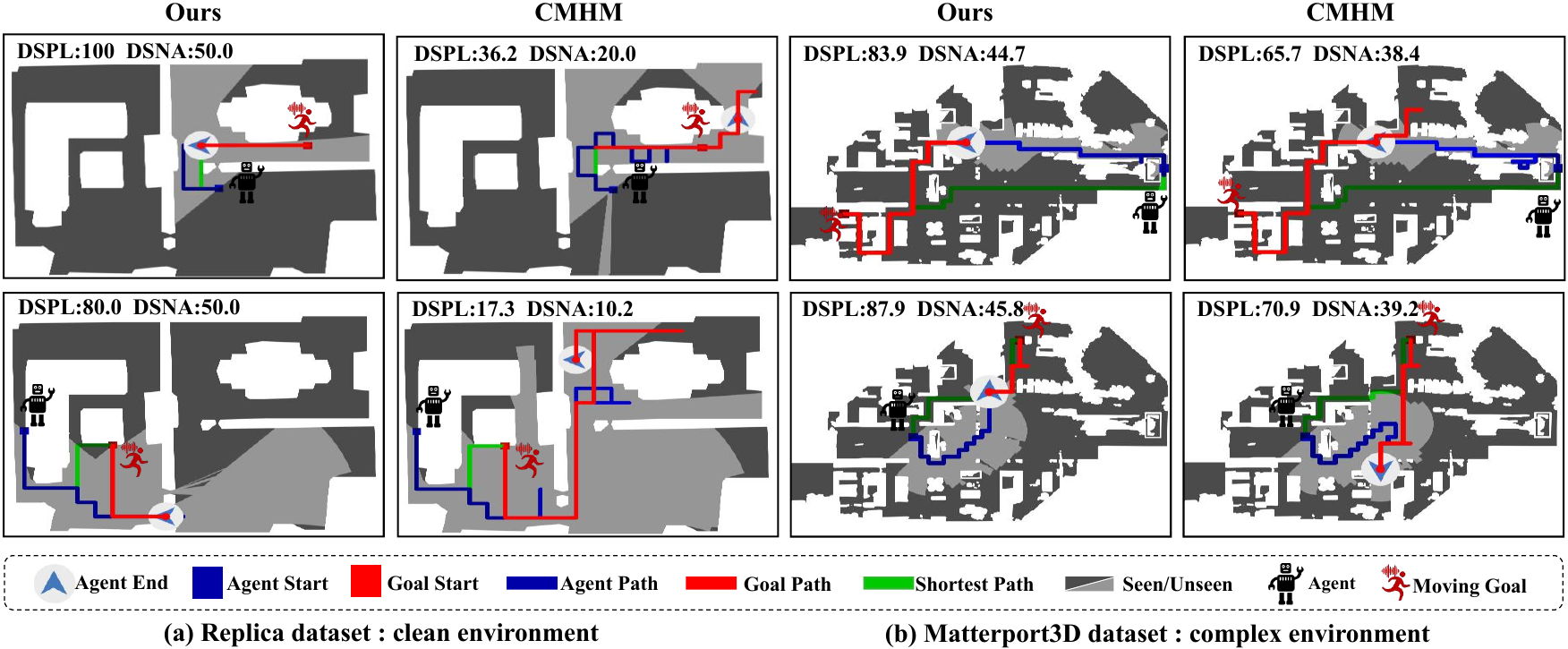}
\caption{Visualization comparison of navigation paths between our method and the strongest baseline across different unheard sound categories.}
\label{fig:figure7}
\end{figure*}

\subsubsection{Qualitative Analysis}
To gain deeper insights into the proposed framework, we conduct a qualitative analysis of its two stages: Physical Alignment and Scene Aggregation.

Figure~\ref{figure3} visualizes physics-informed physical alignment in Stage I. Compared with the pure-visual encoder (Figure~\ref{figure3}(a)), which mainly captures geometric structure, the ICGE (Figure~\ref{figure3}(b)) shows pronounced activations aligned with the sound source direction. This indicates that ICGE effectively calibrates sound intensity onto geometric representations, learning a category-agnostic spatial prior based on propagation-consistent cues (e.g., attenuation and occlusion) rather than source-specific semantics, which underpins the observed $14.4\%$ performance gain on unheard sound categories. This spatial prior is recalibrated online, as $A_t$ records intensity only at visited locations rather than a global sound field.

\begin{table*}[t]
\centering
\caption{Ablation study and computational efficiency on the Replica dataset. Efficiency measured on an A40 GPU (5 parallel processes; batch size 1). Latency excludes simulator; Train FPS includes it.}
\label{tab:ablation_efficiency}
\begin{tabular}{l ccc ccc ccc}
\toprule
\multirow{2}{*}{\textbf{Model Variant}}
&
\multicolumn{3}{c}{\textbf{Heard}}
&
\multicolumn{3}{c}{\textbf{Unheard}}
&
\multicolumn{3}{c}{\textbf{Efficiency}}
\\
\cmidrule(lr){2-4}
\cmidrule(lr){5-7}
\cmidrule(lr){8-10}
&
DSR$\uparrow$
&
DSPL$\uparrow$
&
DSNA$\uparrow$
&
DSR$\uparrow$
&
DSPL$\uparrow$
&
DSNA$\uparrow$
&
Params (M)$\downarrow$
&
Train FPS$\uparrow$
&
Latency (ms)$\downarrow$
\\
\midrule
Baseline1 ($h=512$)
& 90.0 & 62.6 & 34.5
& 49.1 & 29.4 & 16.1
& 13.48 & 35.87 & 3.19 \\

Baseline2 ($h=256$)
& 89.3 & 63.2 & 35.5
& 49.3 & 28.7 & 16.0
& 6.16 & 34.68 & 3.15 \\
\midrule
w/ ICGE
& 91.2 & 66.4 & 36.5
& 54.2 & 30.6 & 17.2
& 6.62 & 33.65 & 4.43 \\

w/ CMSA
& 91.5 & 64.2 & 36.0
& 51.8 & 31.1 & 17.8
& 3.63 & 33.83 & 4.39 \\
\midrule
\textbf{Full Model (Ours)} 
& \textbf{93.8}
& \textbf{67.5}
& \textbf{37.5}
& \textbf{63.5}
& \textbf{32.1}
& \textbf{17.8}
& 3.90
& 32.22
& 5.84 \\
\bottomrule
\end{tabular}
\end{table*}

We further analyze the Scene Aggregation in Stage II by examining the modal utilization patterns of the CMSA. Figure~\ref{figure4} and Figure~\ref{figure5} illustrate the attention distribution across multiple encoders over 10 episodes on Replica and MP3D, respectively. In both datasets, a consistent trend emerges: the model primarily relies on Audio tokens for global guidance, while Visual tokens serve as essential structural supplements. The ICG and SA tokens complement each other, demonstrating CMSA's ability to actively utilize multiple encoders based on scene dynamics to assist decision-making.
Figure~\ref{figure6} provides a detailed view of this behavior by visualizing modal utilization during an MP3D episode. When the agent is far from the target, the ICG and SA tokens receive the highest attention, providing robust spatial priors and directional cues. As the agent approaches the target, attention becomes more balanced across all four modalities (V, A, SA, and ICG), reflecting a transition from coarse localization to fine-grained multi-modal integration. This demonstrates that CMSA dynamically adapts modal reliance to the navigation context.

Finally, Figure~\ref{fig:figure7} illustrates the benefits of the two-stage design. Compared with the strong baseline~\cite{younes2023catch}, our agent generates smoother and more direct trajectories. In the first Replica example, it quickly converges to the moving target, whereas the baseline follows a more circuitous path. These observations are consistent with the improved DSPL and DSNA results, demonstrating the effectiveness of physics-informed alignment and dynamic scene aggregation.

\subsection{Ablation Studies}
To evaluate the individual and combined contributions of ICGE and CMSA, we conducted ablation studies, with the main analysis based on the clean Replica results in Table~\ref{tab:ablation_efficiency}. Comparing Baselines 1 and 2 shows that increasing the hidden dimension provides negligible improvement, indicating that performance is limited by feature fusion rather than GRU capacity. Therefore, Baseline 2 is used in the following analysis.

Integrating ICGE into the baseline (w/ ICGE) significantly boosted performance in unheard categories (DSR $54.2\%$ vs. $49.3\%$). This confirms that physics-informed physical alignment between sound intensity and geometry provides a critical category-agnostic prior, enhancing adaptability to novel sounds without relying on semantic features. Conversely, adding CMSA alone (w/ CMSA) improved DSPL across all settings. This validates that preserving spatial tokens and enabling dynamic Transformer-based interactions yields superior state representations compared to static vector concatenation.

\begin{table}[ht]
\centering
\caption{Ablation Study on Layer Number on Replica Dataset.}
\label{tab:layer_ablation_simplified}
\begin{tabular}{l ccc} 
\toprule
\textbf{Model Variant} & \multicolumn{3}{c}{\textbf{Heard}} \\
\cmidrule(lr){2-4} 
& DSR~$\uparrow$ & DSPL~$\uparrow$ & DSNA~$\uparrow$ \\
\midrule
Baseline & 90.0 & 62.6 & 34.5 \\
Variant-L1 & 87.6 & 62.7 & 36.3 \\
Variant-L2 & 90.8 & 63.0 & 36.9 \\
Variant-L3 & 91.6 & 65.1 & 35.8 \\
\midrule
\textbf{Full Model (Ours)} & \textbf{93.8} & \textbf{67.5} & \textbf{37.5} \\
\bottomrule
\end{tabular}
\end{table}

The Full Model achieved optimal performance across all metrics, reaching a DSR of $63.5\%$ in unheard categories. 
This demonstrates a powerful synergy: ICGE provides physics-informed, generalizable features, while CMSA enables intelligent, fine-grained fusion of those features. Additionally, varying the Transformer layers (Table~\ref{tab:layer_ablation_simplified}) showed that a 4-layer depth provides sufficient capacity for deep cross-modal reasoning, outperforming shallower configurations.

\subsection{Limitation Analysis}
Despite its improvements, our method still has several limitations.
First, performance gains diminish in large-scale environments: the DSR improvement is $14.4\%$ on Replica but decreases to $11.7\%$ on MP3D, suggesting that the benefits of physics-informed alignment are partly offset by increased search space and topological complexity.
Second, although the multi-layer Transformer in CMSA enables fine-grained cross-modal interaction, it increases attention computation (Table~\ref{tab:ablation_efficiency}). The full model reduces parameters from $6.16$M to $3.90$M ($\sim$37\%) relative to Baseline2, yet the model-side per-frame latency (excluding simulator interaction) rises from $3.15$ ms to $5.84$ ms, which may limit real-time deployment on resource-constrained edge devices.
Finally, the relatively small DSPL and DSNA gains on MP3D indicate limited search efficiency. Without explicit long-term memory, the agent may perform redundant exploration in multi-loop environments, suggesting that external memory is a promising direction for improving large-scale navigation.

\section{Conclusion}
We address D-AVN’s cross-category generalization limitation stemming from overreliance on sound semantics with a two-stage multimodal framework. ICGE conducts physics-informed spatial alignment via intensity-geometry fusion, and the CMSA Transformer achieves fine-grained dynamic multimodal aggregation. Tests on Replica and Matterport3D outperform SOTA, excelling on unseen sound categories and complex acoustic environments. Ablations validate that the two modules jointly enable stable, efficient navigation and improve embodied agents’ generalization.

\section*{Acknowledgments}
This research was financially supported by the National Natural Science Foundation of China (Grant No.: 62463029) and the Natural Science Foundation of Xinjiang Uygur Autonomous Region of China (Grant No.: 2026D01C56).

\bibliographystyle{IEEEtran}
\bibliography{ref}

\end{document}